\documentclass[1p,times]{elsarticle}
\usepackage{amsmath,amssymb}
\usepackage[colorlinks]{hyperref}
\usepackage[hyperref]{xcolor}
\usepackage{natbib}
\usepackage{pxfonts}
\usepackage{tabularx}
\usepackage{tikz}
\usetikzlibrary{calc,intersections,through,hobby}
\usepackage{cleveref}
\newcommand{\CL}[1]{\ensuremath{\mathrm{Cl_{#1}}}}
\newcommand{\Li}[1]{\ensuremath{\mathrm{Li}_{#1}}}
\newcommand{\ZN}{\ensuremath{\mathcal{Z}}}
\begin{document}
\title{Critical behavior of isotropic systems with strong dipole-dipole interaction: three-loop study}
\author[label1]{A.\,Kudlis}
\ead{andrew.kudlis@metalab.ifmo.ru} 
\author[label2]{A.\,Pikelner}
\ead{pikelner@theor.jinr.ru} 
\address[label1]{Faculty of Physics, ITMO University, Kronverkskiy prospekt 49, St. Petersburg 197101, Russia}
\address[label2]{Bogoliubov Laboratory of Theoretical Physics, Joint Institute for Nuclear Research, 6 Joliot-Curie, Dubna, 141980, Russia}

\begin{abstract}
  We analyze the critical behavior of isotropic systems with dipole-dipole
  interaction by renormalization-group methods in fixed space-time dimensions.
  Working in three-dimensional theory we analytically find three-loop expressions
  for critical exponents in the limit of dominating dipole-dipole forces.
  Resummation of the series obtained provides numerical values close to
  $O(3)$-theory predictions, justifying the applicability of such a simplified
  model to systems with strong dipole-dipole interaction.
\end{abstract}
\maketitle

\section{Introduction}
\label{sec:intr}
The role of the $O(n)$-symmetric model is difficult to overestimate in studying
the critical behavior of a wide range of physical systems. To process this model
correctly, Wilson proposed the elegant idea to apply the renormalization group
(RG) approach and developed the $\varepsilon$ expansion technique as a
computational
procedure~\cite{PhysRevB.4.3174,PhysRevB.4.3184,WILSON1974119,PhysRevLett.28.240}.
The five-loop calculation performed almost forty years ago was considered as a
record
one~\cite{CHETYRKIN1983351,CHETYRKIN1981147,LeGuillou1985,KLEINERT199139}, and
the situation remained unchanged for a long time.

Due to the impressive development of numerical
methods~\cite{BinothHeinrich:SectorDecomposition,Baikov2010mi,Lee2012mi,chetyrkin2017rstar,Batkovich2015rstar,Brown:TwoPoint,Panzer:HyperIntAlgorithms,BrownKreimer:AnglesScales,BatkovichKompanietsChetyrkin:6loop,KompanietsPanzer:LL2016},
as well as computational power of modern hardware, the six- and seven-loop
results in $\varepsilon$ were obtained in Refs.~\cite{KP17}
and~\cite{PhysRevD.97.085018}, respectively. The found values of the diagrams
allowed to analyze the critical behaviour in the following orders of
perturbation theory in the case of different $O(n)$-symmetric generalizations,
which specify symmetries of various physical
systems~\cite{six_loop_cubic,six_loop_onom,six_loop_unum,Bednyakov2021}.
 
As known, there are various RG approaches. In addition to the $\varepsilon$
expansion, it is possible to construct a theory directly in physical spatial
dimensions. To this day, for the three-dimensional (3d) $O(n)$-symmetric model
the record -- six-loop -- calculation was performed by Nickel et al. in
Ref.~\cite{NickelREPORT,NickelPRB1976}. That made it possible to find the RG
functions and accurate values of the critical
exponents~\cite{PhysRevB.21.3976,PhysRevLett.39.95}.

The real-space RG can be advantageous when the action has a specific term. For
example, the three-dimensional approach is extremely natural when the action
contains a term that requires the coincidence of order parameter and spatial
dimensions. In particular, taking into account dipole-dipole interaction leads
to the model containing such a term. We should note that this model has only
been studied in the low-order approximation within the $\varepsilon$
expansion~\cite{PhysRevLett.30.559,PhysRevB.10.2078} as well as with the help of
RG in physical spatial
dimension~\cite{Korzhenevskii1978,9168923975ca4bcebb5c0a2bc92ab3b5}. Since
calculations for the systems with dipole-dipole interaction is quite
complicated, but for a number of physical systems the inclusion of this
interaction is
important~\cite{KORNBLIT1973531,RevModPhys.46.597,RevModPhys.52.489,WACHTER1979507},
we aim to estimate this effect by considering a simplified model that is a
limiting case of complete dominance of dipole forces. To do this, we obtain the
three-loop RG series, on their basis we extract the values of critical
exponents. Comparison of the obtained numbers with their counterparts in the
case of the standard $O(n)$-symmetric universality class
from~\cite{PhysRevB.21.3976} allows us to evaluate the need to consider the term
responsible for dipole-dipole interaction in the analysis of critical behavior.

As a starting point, we reproduce the results for the simple $O(n)$-symmetric
model (SSM), which is presented in
Ref.~\cite{NickelPRB1976,NickelREPORT,PhysRevB.17.1365}. We managed to get
answers up to three loops completely analytically. The technique for calculating
diagrammatic series is further applied to the case of the model with strong
dipole-dipole interaction (SDM). All the RG expansions and final numerical
estimates of observables are presented both for SSM and SDM.

The paper is organized as follows. In Sec.~\ref{sec:model_and_ren}, the model
and the renormalization scheme, which we use in this work, are described. After
that, the calculation details, including explanation of diagrammatic technique,
are presented in Sec.~\ref{sec:calc-details}. Next, in Sec.~\ref{sec:num_res},
the expansions for RG functions and critical exponents as well as their
numerical estimates are shown. Finally, in Sec.~\ref{sec:conclusion} we draw a
conclusion.

\section{Model and renormalization scheme}
\label{sec:model_and_ren}
Thus, we are interested to analyze the critical behaviour of the following model: 
\begin{equation}
  S = - \frac{1}{2} \varphi_{\alpha} \left(p^2 + m_0^2\right)\Omega_{\alpha\beta}\varphi_{\beta}
  -\frac{1}{4!} \lambda T^{\alpha\beta\gamma\delta}\varphi_{\alpha}\varphi_{\beta}\varphi_{\gamma}\varphi_{\delta},\label{eqn:our_model}
\end{equation}
where $\varphi_{\alpha}$ is $d$-component bare field, $\lambda$ is bare coupling
constant, $m_0^2$ is bare mass being proportional to $T-T_c$, where $T_c$ is the
mean-field critical temperature. Tensor $\Omega_{\alpha\beta}$ is defined as
follows:
\begin{equation}
  \textrm{SSM:\ } \Omega_{\alpha\beta} = \delta_{\alpha\beta},
  \quad
  \textrm{SDM:\ } \Omega_{\alpha\beta}=P^T_{\alpha\beta}=\left(\delta_{\alpha\beta}-\frac{p_{\alpha}p_{\beta}}{p^2}\right).
\end{equation}
The modification of tensor $\Omega_{\alpha\beta}$ in the case of SDM arises from
the Fourier transform of dipolar interaction term
$\sim(x^2\delta_{\alpha\beta}-x_{\alpha}x_{\beta})/x^5$. Tensor factor
$T^{\alpha\beta\gamma\delta}$ reads as:
\begin{equation}
  T^{\alpha\beta\gamma\delta} = \frac{1}{3}(\delta_{\alpha \beta} \delta_{\gamma \delta} +\delta_{\alpha \gamma} \delta_{\beta \delta}+\delta_{\alpha \delta} \delta_{\gamma \beta})
\end{equation}
This model is studied directly in $d=3$. The propagator of the theory can be
written in the following form:
\begin{equation}
  G_0^{\alpha\beta}(p)=\Omega_{\alpha\beta}\dfrac{1}{p^2+m_0^2}.\label{eq:GT-def}
\end{equation}

Our calculation scheme is based on the one used
in~\cite{NickelREPORT}\footnote{Available for download from
\url{http://users.physik.fu-berlin.de/~kleinert/nickel/guelph.pdf}} and can be
explained on the examples from the SSM model, results of calculation are in one
to one correspondence with the three-loop results presented
in~\cite{NickelREPORT}. For the SDM model we proceed in a similar way up to the
redefinition of $\Omega_{\alpha\beta}$.

Our final goal is to determine renormalization constants $Z_i(u)$ as a function
of the renormalized dimensionless coupling $u$. However, for actual diagrams
calculation it is convenient to introduce an intermediate scheme, where the
propagator is defined by~\eqref{eq:GT-def} and perturbative series are organized
as expansion in auxiliary parameter $W$. For dimensionful quantities, such as
four-point function in $d=3$, result of calculation depends on $m_0$ and may
diverge due to divergent subdiagrams. Such divergences are removed by a
specially constructed mass counter-term and after eliminating bare mass with
$m_0^2=Z_3^{-1}m^2-\delta m_0^2$ the final result becomes finite. The exact form
of $\delta m^2$ can be determined from the mass derivative of the two-point
function and depends on the chosen regularisation prescription. In original
work~\cite{NickelREPORT} divergent subgraphs were subtracted under the integral
sign making sum of the diagram and diagram with counter-term insertion finite.
In our work, we make use of dimensional regularization, thus both diagram and
its counter-part now have poles in dimensional regularization parameter
$\varepsilon$, but the sum is finite and the same as results presented
in~\cite{NickelREPORT}.

The specific choice of connection $m_0^2=Z_3^{-1}m^2-\delta m_0^2$ between bare
and renormalized mass greatly simplifies calculation allowing independent
extraction of renormalization constants in the intermediate scheme as series in
parameter $W= - \sqrt{Z_3}\lambda/m$ from particular diagrams. Similar to
work~\cite{NickelREPORT}, we are interested in the set of renormalization
constants $\ZN_i(W)$ in the intermediate scheme defined as
follows\footnote{Numerical coefficients of the $\ZN_i$ expansion in $W$ for the
SSM model are exactly the same as numbers provided in~\cite{NickelREPORT}, Table
5.}:
\begin{align}
  \ZN_1^{-1}(W) & =\left.
                  -\frac{1}{\lambda}\Gamma^{(4,0)}\left(p,\lambda,m^2Z_3^{-1}\right)\right|_{p^2=0},
                  \label{eq:ZN1}\\
  \ZN_4^{-1}(W) & =\left.
                  \Gamma^{(2,1)}\left(p,\lambda,m^2Z_3^{-1}\right)\right|_{p^2=0},
                  \label{eq:ZN4}\\
  \ZN_3^{-1}(W) & = \frac{\Omega_{\alpha\beta}}{\Omega^2}\left.\left(\frac{\partial}{\partial p^2} \Gamma^{(2,0)}_{\alpha\beta} \left(p,\lambda,m^2Z_3^{-1}\right)\right)\right|_{p^2=0}.
                  \label{eq:ZN3}
\end{align}
Here, $\Gamma^{(i,j)}$ is $i$-point function with $j$ insertions of operator
$O_2 = \Omega_{\alpha\beta}\varphi_{\alpha}\varphi_{\beta}$. Especially, in our
case we calculate the four-point function with zero external momenta
in~\eqref{eq:ZN1}, two-point function with $O_2$ insertion and zero external
momenta in~\eqref{eq:ZN4} and zero external momentum limit of the momentum
derivative of the two-point function with the appropriate projector applied
in~\eqref{eq:ZN3}.

Results for renormalization constants as functions of the renormalized
dimensionless coupling $Z_i(u)$ can be derived
from~\eqref{eq:ZN1},~\eqref{eq:ZN4}, and~\eqref{eq:ZN3} with:
\begin{equation}
  \label{eq:ZuFromZw}
  Z_i(u) = \ZN_i\left(\frac{ - u Z_1(u)}{Z_3^{3/2}(u)}\right), \quad i={1,3,4},
\end{equation}
where we used the relation $\lambda=u m Z_1/Z_3^2$, which binds the bare and
renormalized couplings.

Having obtained all the necessary $Z_i(u)$, we can derive the expressions for
all the RG functions. To obtain the series for $\beta$-function, we can utilize
the following expression:
\begin{equation}
  \beta(u)=(d-4)\left(\dfrac{\partial \ln \left[uZ_1(u)/Z_3^2(u) \right]}{\partial u}\right)^{-1}.
\end{equation}
The anomalous dimensions are connected with the renormalization constants in the following way:
\begin{align}
  \gamma_{3}(u) & = \beta(u)\dfrac{\partial \ln\left[Z_3(u)\right]}{\partial u},\\
  \gamma_{4}(u) & = \beta(u)\dfrac{\partial \ln\left[Z_4(u)\right]}{\partial u}.
\end{align}
As well known, the behavior of the system in the vicinity of the critical
temperature is determined by infrared-stable fixed point $u^*$, which is
determined by zero of the $\beta$-function ($\beta(u^*)=0$). Also, it is
convenient to introduce other functions of $u$, in terms of which one could
express the critical exponents:
\begin{equation}
  \eta(u)=\gamma_{3}(u), \quad \nu(u)=\dfrac{1}{2+\gamma_{4}(u) -
  \gamma_{3}(u)}, \quad \gamma(u)=\dfrac{2 - \gamma_{3}(u)}{2 + \gamma_{4}(u) - \gamma_{3}(u)}.\label{eqn:add_functions}
\end{equation}
We limit ourselves by consideration of the most traditional critical exponents
$\eta$, $\nu$, and $\gamma$, which are defined by functions
from~\eqref{eqn:add_functions} evaluated at $u=u^*$: $\eta=\eta(u^*)$,
$\nu=\nu(u^*)$, and $\gamma=\gamma(u^*)$. In addition, it is necessary to define
the leading corrections to the scaling law, which is governed by the following
exponent:
\begin{equation}
  \omega=\omega(u^*)=\left.\dfrac{\partial \beta_{u}}{\partial u}\right\rvert_{u=u^*}.
\end{equation}
However, before we turn to the numerical results, let us tell a few words about
the technique for calculating the diagrammatic expansions.

\section{Details of calculation}
\label{sec:calc-details}

As was mentioned in Sec.~\ref{sec:model_and_ren}, we calculate individual
diagrams in the framework of dimensional regularization as expansion in
parameter $\varepsilon = 3-d$. Compared to a more traditional numerical
evaluation technique used in~\cite{NickelREPORT}, the chosen approach at the
first sight seems overcomplicated, but it has several attractive features. First
of all, working at the three-loop level we are able to perform all calculations
analytically, which is especially useful for accurate divergences subtraction
and manipulation with the obtained expressions. Secondly, the dimensional
regularization framework allows to apply efficient tools for modern multi-loop
calculations. Also, the availability of three-loop results for dimensionally
regulated massive tadpole~\cite{Rajantie:1996np} integrals justifies our choice
of the regularization scheme.

In the three-loop calculation, presented in the paper, a number of diagrams to
be considered is not so high, especially compared to six-loop
calculation~\cite{NickelREPORT}. However, we decided to develop a highly
automized setup to simplify tensor manipulations, especially in the case of SDM,
and to eliminate possible errors. One more complication comes from the
transverse structure of the propagator in SDM~(\ref{eq:GT-def}) leading to the
appearance of new classes of diagrams with massless propagators. Calculation of
such diagrams cannot be reduced to the set of integrals provided
in~\cite{NickelREPORT} and they need a separate treatment.

We start our chain of calculations with \texttt{DIANA}~\cite{Tentyukov:1999is},
which internally calls \texttt{QGRAF}~\cite{Nogueira:1991ex} to generate
diagrams in both models through the single run. After substitution of Feynman
rules for a specific model and application of appropriate projectors, we perform
partial fraction decomposition and map the obtained scalar integrals on
predefined set of topologies. Inside each topology all integrals are reduced to
a small number of master integrals during the reduction step performed with the
Laporta algorithm implemented in package \texttt{FIRE}~\cite{Smirnov:2019qkx}
with the inclusion of additional symmetry rules between integrals generated with
package \texttt{LiteRed}~\cite{Lee:2012cn}.

After reduction to master integrals, in order to obtain results of the
individual diagrams as the series in $\varepsilon$, we substitute the results of
the $\varepsilon$ expansions for master integrals. It is important to note, that
minimal set of master integrals obtained after reduction step does not
necessarily match original set of diagrams. Some of needed master integrals may
contain poles in $\varepsilon$, or vice versa have divergent coefficients. In
the latter case we need to provide higher orders of expansion in $\varepsilon$
of these master integrals. For all the most complicated integrals we need only
finite parts available
from~\cite{Rajantie:1996np,Broadhurst:1998iq,Broadhurst:1998ke} and for some
higher order expansion results can be found
in~\cite{Lee:2010hs,Davydychev:2000na}.

Results for individual diagrams contributions both in the SSM and SDM can be
found in~\ref{sec:dias-2pt} for the two-point functions and
in~\ref{sec:dias-4pt} for the four-point function and two-point function with
the operator insertion. Present tables contain contribution to the expansion in
variable $W$ and explicitly reads:
\begin{align}
  \ZN_{1,X}^{-1}(W) & = 1
                      + W \Gamma_{X,2}^{(4,0)}
                      + W^2 \sum\limits_{i=3}^{4}\Gamma_{X,i}^{(4,0)}
                      + W^3 \sum\limits_{i=5}^{12}\Gamma_{X,i}^{(4,0)}
                      + \mathcal{O}\left(W^4\right),\label{eq:Z1inv-Nickel}\\
  \ZN_{4,X}^{-1}(W) & = 1
                      + W \Gamma_{X,2}^{(2,1)}
                      + W^2\sum\limits_{i=3}^{4}\Gamma_{X,i}^{(2,1)}
                      + W^3 \sum\limits_{i=5}^{10}\Gamma_{X,i}^{(2,1)}
                      + \mathcal{O}\left(W^4\right),\label{eq:Z4inv-Nickel}\\
  \ZN_{3,X}^{-1}(W) & = 1
                      + W^2 \Gamma_{X,2}^{(2,0)}
                      + W^3 \Gamma_{X,3}^{(2,0)}
                      + \mathcal{O}\left(W^4\right).\label{eq:Z3inv-Nickel}                      
\end{align}
where $X=O$ for the case of SSM and $X=D$ for the case of SDM.

All presented results are finite, since for the divergent four-point diagram
(no.7 in \ref{sec:dias-4pt}) we present the result with the subdivergence
subtracted. For SSM the contributions to $\ZN_1$ from~\eqref{eq:dia7-4-N} and to
$\ZN_4$ from~\eqref{eq:dia7-21-N} coincide with~\cite{NickelREPORT}, where a
different subtraction scheme was utilized.

One more check on the validity of provided results comes from the fact that at
least at the two-loop level we can consider a more general type of diagrams
depending on additional parameter $g$ responsible for dipole-dipole interaction.
The two limiting cases, $g\rightarrow 0$ and $g\rightarrow \infty$, correspond
to the result in SSM and SDM, respectively. As an example for the diagram
covering both $\Gamma_{O,2}^{(4,0)}$ in~\eqref{eq:dia2-4-N} and
$\Gamma_{D,2}^{(4,0)}$ in~\eqref{eq:dia2-4-D} we provide the following
closed-form expression:
\begin{equation}
  \Gamma_{X,2}^{(4,0)}=\frac{102 g+181 \sqrt{g+1}+149}{15 \left(g+\sqrt{g+1}+1\right)}.
\end{equation}

\section{Numerical results}
\label{sec:num_res}
Following pioneering works~\cite{NickelPRB1976,PhysRevLett.39.95}, we modify the
normalization of the coupling constant as $v=11u$, which is dictated by the
one-loop coefficient of $\beta$-function in the case of SSM when $n=3$
($O(3)$-symmetric model). Moreover, in order to compare the results in the
ordinary (labeled as $A_O$) and dipole-dipole (labeled as $A_D$) cases, at all
stages we present the numerical results for both of them. All the expansions are
found analytically. Due to their bulkiness, they are presented
in~\ref{app:analyt_rg_expansions}. Here, we restrict ourselves to the results
through decimal notation. The number of decimal places was chosen to be the same
as for Nickel et al. in work~\cite{PhysRevB.17.1365} despite the fact that they
are known with absolute accuracy. Thus, the three-loop expansions for the RG
functions have the following form:
\begin{align}
  \beta_O(v) & =-v + 1.00000000000 v^2 - 0.3832262014 v^3 + 0.2829466816 v^4  + \mathcal{O}(v^5),\\ 
  \beta_D(v) & =-v + 0.61818181812 v^2 - 0.1036254517 v^3 + 0.0416046792 v^4  + \mathcal{O}(v^5),\\ 
  \eta_O(v) & = 0.0122436486 v^2 + 0.00102000004 v^3  + \mathcal{O}(v^4),\\
  \eta_D(v) & = 0.0053650924 v^2 + 0.00074891316 v^3  + \mathcal{O}(v^4),\\
  \nu_O(v) & = 0.5 + 0.1136363636 v + 0.0082262014 v^2 +  0.0112625247 v^3 + \mathcal{O}(v^4),\\
  \nu_D(v) & = 0.5 + 0.0757575758 v + 0.0063917781 v^2 +  0.0025065562 v^3 + \mathcal{O}(v^4),\\
  \gamma_O(v) & = 1.0 + 0.2272727274 v + 0.0103305785 v^2 + 0.0206237256 v^3 + \mathcal{O}(v^4),\\
  \gamma_D(v) & = 1.0 + 0.1515151515 v + 0.0101010101 v^2 + 0.0042322094 v^3 + \mathcal{O}(v^4),\\
  \omega_O(v) & = -1.0 + 2.0000000000 v - 1.1496786042 v^2 + 1.1317867264 v^3 + \mathcal{O}(v^4),\\
  \omega_D(v) & = -1.0 + 1.2363636364 v - 0.3108763550 v^2 + 0.1664187168 v^3 + \mathcal{O}(v^4).
\end{align}
\begin{table}[t!]
  \begin{center}
    \caption{Comparison of numerical estimates of fixed point coordinates for SSM
      and SDM within the same normalization. The numbers were extracted based on the
      $\tau$-expansions~\eqref{eqn:tau_expansions_for_coordinates}. The final answer
      is given by averaging the Pad\'e and PB numerical estimates. The difference
      between the six-loop answer and three-loop result for the usual symmetric
      model without dipole-dipole interaction is taken as an error. This estimate is
      extrapolated to the case of the dipole-dipole model. The results of direct
      summation are shown here only to demonstrate the convergence of the
      series.}
    \label{tab:fixed_points}
    \setlength{\tabcolsep}{14.6pt}
    \renewcommand{\arraystretch}{1.4}
    \begin{tabular}{{c}*{4}{c}}
      \hline
      \hline
      Coordinate & Direct Summation & Pad\'e[2/1] &  PB[2/1] & Final \\
      \hline
      $v_O^*$ & $1.39400$ &  $1.39432$ & $1.39443$  & $1.394(4)$\\
      $v_D^*$ & $2.00930$ &  $2.01385$ & $2.01502$  & $2.014(6)$\\
      \hline
      \hline
    \end{tabular}
  \end{center}
\end{table}

Let us extract now the numerical values of the critical exponents. For this, as
already mentioned, we need to find the value of the coordinate of fixed point
$v^*$. Due to the asymptotic nature of renormalization group series, in order to
obtain the proper numerical results the various resummation techniques should be
applied. The most basic of them is the method of Pad\'e approximants. However,
in such an approximation, this method applied to the initial RG expansion does
not give any reliable results -- all the Pad\'e approximants give drastically
different values. The Pad\'e-Borel (PB) method applied to the original RG series
does not improve the situation considerably. However, this dramatic picture can
be improved in the following way. B. Nickel (see Ref. [19] in
Ref.~\cite{PhysRevB.21.3976}) proposed a method to advance the convergence of
series. This method consists of reexpanding of the renormalization group series
into alternative power expansions whose coefficients demonstrate a more
favorable behavior. The idea is to put a formally small parameter $\tau$ at the
linear term of the $\beta$-function ($\beta(v)=-\tau v+ v^2 + \dots$). Then,
similarly to how it is done within the $\varepsilon$ expansion, one can
iteratively find a fixed point coordinate as series in $\tau$. After further
processing with various resummation techniques of new expansion, which already
has a more acceptable structure, the formal parameter $\tau$ should be equated
to unity. Following this recipe for fixed point coordinates in both cases we
obtain:
\begin{align}\label{eqn:tau_expansions_for_coordinates}
  v_O^* & = 1.0000000000 \tau + 0.3832262014  \tau^2 + 0.0107779613 \tau^3 + \mathcal{O}(\tau^4), \\
  v_D^* & = 1.6176470588 \tau + 0.4386496164  \tau^2 - 0.0469970036 \tau^3 + \mathcal{O}(\tau^4).
\end{align}
Based on these expansions we extract the numerical estimates for coordinates
$v_O^*$ and $v_D^*$ by means of naive direct summation, Pad\'e approximants, as
well as PB technique. These numbers are presented in
Table~\ref{tab:fixed_points}. We choose diagonal approximant $[2/1]$ for the
Pad\'e and PB estimates, given that approximants should be constructed for
series starting with a constant. Noteworthy, as a test, we have the six-loop
fixed-point estimate in the SSM case~\cite{PhysRevLett.39.95,KUDLIS2020114881}:
$1.391(1)$. Believing in the convergence of estimates with increasing orders of
perturbation theory, we can choose its difference from the six-loop value as an
error for the final three-loop estimate: $1.394(4)$. Following this logic, we
can try to interpolate the calculation error per estimate in the case of a model
with a dipole-dipole interaction: $2.014(6)$.

Let us move on to the analysis of the most interesting quantities from the
physical point of view - the critical exponents. The situation with exponents is
similar. The analysis of the initial RG series leaves much to be desired. To
extract some proper estimates, we reexpand the RG expansions for functions
$\eta(v)$, $\nu(v)$, $\gamma(v)$, and $\omega(v)$ in terms of $\tau$. They read
as follows:
\begin{align}
  \eta_O(\tau) & = 0.0122436486 \tau^2 + 0.0104041739  \tau^3  + \mathcal{O}(\tau^4),\\
  \eta_D(\tau) & = 0.0140392772 \tau^2 + 0.0107840990 \tau^3  + \mathcal{O}(\tau^4),\\
  \nu_O(\tau) & = 0.5 + 0.1136363636 \tau + 0.0517746334 \tau^2 +  0.0187922848 \tau^3 + \mathcal{O}(v^4),\\
  \nu_D(\tau) & = 0.5 + 0.1225490196 \tau + 0.0499569216 \tau^2 +  0.0161209071 \tau^3 + \mathcal{O}(\tau^4),\\
  \gamma_O(\tau) & = 1.0 + 0.2272727274 \tau + 0.0974274425 \tau^2 + 0.0309911590 \tau^3 + \mathcal{O}(\tau^4),\\
  \gamma_D(\tau) & = 1.0 + 0.2450980392 \tau + 0.0928942046 \tau^2 + 0.0251292650 \tau^3 + \mathcal{O}(\tau^4),\\
  \omega_O(\tau) & = -1.0 + 2.0000000000 \tau - 0.3832262014 \tau^2 + 0.2721687203 \tau^3 + \mathcal{O}(\tau^4),\\
  \omega_D(\tau) & = -1.0 + 2.0000000000 \tau - 0.2711652174 \tau^2 + 0.2051665366 \tau^3 + \mathcal{O}(\tau^4).
\end{align}
Once we have the truncated series, we can apply to them the summation
techniques, which were used previously for the fixed point coordinates. The
corresponding numbers are presented in Table~\ref{tab:critical_exponents}. As
previously, believing in the succession of the model with the strong
dipole-dipole interaction compared to the conventional symmetric model, to
estimate the final values of the errors, we resorted to the results obtained
within the six-loop calculation of the symmetric theory only with the exchange
interaction~\cite{PhysRevLett.39.95} ($\nu=0.7054(11)$, $\eta=0.0340(25)$,
$\gamma=1.3866(12)$, and $\omega=0.779(6)$).
\begin{table}[t!]
  \begin{center}
    \caption{Comparison of numerical estimates of critical exponents for SSM and
      SDM. The numbers were extracted based on the $\tau$-series. Pad\'e and PB
      estimates are obtained according to the near-diagonal 3-loop values ($[2/1]$ and
      $[1/2]$ since the diagonal is absent in this case) and according to the diagonal
      two-loop $[1/1]$. The average value based on the Pad\'e and PB estimates is
      given as an answer. The dash corresponds to the fact that some of the
      approximants in the sample are spoiled by the presence of a dangerous pole,
      which does not allow to make an estimate.}
    \label{tab:critical_exponents}
    \setlength{\tabcolsep}{15.6pt}
    \renewcommand{\arraystretch}{1.3}
    \begin{tabular}{{c}*{4}{c}}
      \hline
      \hline
      Exponent & Direct Summation & Pad\'e &  PB & Final \\
      \hline
      $\eta_O$ & $0.02265$ &  $0.03880$ & $-$  & $0.039(9)$\\
      $\eta_D$ & $0.02482$ &  $0.03314$ & $-$  & $0.033(8)$\\
      \hline
      $\gamma_O$ & $1.35569$ &  $1.38024$ & $-$  & $1.380(8)$\\
      $\gamma_D$ & $1.36312$ &  $1.38093$ & $-$  & $1.381(8)$\\
      \hline
      $\nu_O$ & $0.68420$ &  $0.69986$ & $-$  & $0.700(7)$\\
      $\nu_D$ & $0.68863$ &  $0.70019$ & $-$  & $0.700(7)$\\
      \hline
      $\omega_O$ & $0.88894$ &  $0.76505$ & $0.75968$  & $0.762(17)$\\
      $\omega_D$ & $0.93400$ &  $0.83878$ & $0.84737$  & $0.843(19)$\\
      \hline
      \hline
    \end{tabular}
  \end{center}
\end{table}
Based on the results presented in Table~\ref{tab:critical_exponents}, we can
conclude that the strong dipole-dipole interaction ($g\rightarrow\infty$) does
not strongly change the universality class regarding the critical exponents.
Only the scaling correction exponent has undergone significant changes. Such
numerical results, assuming that we will not find any surprises in higher
orders, to some extent justify the constantly neglected term in the action,
which is responsible for the dipole-dipole interaction.

\section{Conclusion}
\label{sec:conclusion}
In this work, we have calculated the three-loop RG expansions for the isotropic
field model in the limiting dipole-dipole region within the renormalization
group in the physical spatial dimensionality ($d=3$). Such a regime is extremely
important to take into account when analyzing the critical behavior in a number
of ferromagnets and ferroelectrics. All the results have been found completely
analytically, which is of particular value for this area. At each step of our
calculations, all computational procedures were performed both for the model
with dipole-dipole interaction (SDM) and for the usual isotropic model (SSM).
The obtained values for the critical exponents, which are the main measure of
the difference between two different universality classes, turn out to be very
close between the two models -- SSM and SDM, except for the correction for the
scaling exponent $\omega$. To some extent, this justifies the fact that, when
analyzing the critical behavior of such systems, the term responsible for the
dipole-dipole interaction is often neglected in the action. The results of the
diagrams calculation obtained here can be applied to the cases of different
symmetries, and the techniques used here can break into higher orders of
perturbation theory.

\section*{Acknowledgement}
We are deeply grateful to L.~Ts. Adzhemyan for checking some of the
calculations, as well as for lengthy discussions regarding the present problem.
We are also grateful to N. Lebedev for a careful reading of the manuscript and
for fruitful discussion. The work of A.K. was supported by Grant of the Russian
Science Foundation No 21-72-00108.

\appendix

\section{Analytical expansions of RG functions}
\label{app:analyt_rg_expansions}
Here, we present the expansions for the minimal set of RG functions, others can
be extracted on their basis. The three-loop $\beta$-functions for SSM and SDM:
\begingroup
\allowdisplaybreaks
\begin{align}
  \beta_{D}(v) = & -v+\frac{34}{55}v^2-\Bigg[\frac{19744}{81675}-\frac{64 \ln (2)}{165}+\frac{72 \ln (3)}{605}\Bigg]v^3+ \Bigg[\frac{86 \sqrt{2} \text{Cl}_2(4\alpha)}{6655}-\frac{86\sqrt{2}\text{Cl}_2(2 \alpha)}{6655}\nonumber\\
                 &+\frac{411\sqrt{2}\arccos(3)^2}{6655}-\frac{1644\sqrt{2} \text{Li}_2\left(3-2 \sqrt{2}\right)}{6655}-\frac{559851 \text{Li}_2\left(\frac{1}{3}\right)}{106480}+\frac{10741 \pi^2}{38720}\nonumber\\
                 &+\frac{274\pi^2 \sqrt{2}}{6655}-\frac{6673790261}{5995489500}-\frac{22059 \ln^2(3)}{9680}+\frac{7096530443 \ln(2)}{1199097900}\nonumber\\
                 &+\frac{15808 \ln(2) \ln(3)}{6655}-\frac{3339183649 \ln(3)}{1199097900}\Bigg]v^4 +\mathcal{O}(v^5),\\
  \beta_{O}(v) = & -v+v^2-\frac{1252}{3267}v^3+\Bigg[\frac{3552\sqrt{2} \text{Cl}_2(4\alpha)}{1331}-\frac{3552\sqrt{2}\text{Cl}_2(2 \alpha)}{1331}+\frac{23168\text{Li}_2\left(\frac{1}{3}\right)}{1331}\nonumber\\
                 &-\frac{2896\pi^2}{3993}-\frac{87542}{35937}+\frac{5792\ln^2(3)}{1331}+\frac{77684\ln(3)}{11979}-\frac{155368 \ln(2)}{11979}\Bigg]v^4+\mathcal{O}(v^5).
\end{align}
The RG functions $\nu(v)$ and $\gamma(v)$ read as:
\begin{align}
  \nu_D(v) = & \frac{1}{2}
               +\frac{5}{66}v
               +\Bigg[\frac{8 \ln (2)}{165}-\frac{197}{32670}-\frac{7 \ln (3)}{363} \Bigg]v^2
               +\Bigg[\frac{113557\pi^2}{10222080}-\frac{141419\text{Li}_2\left(\frac{1}{3}\right)}{851840}\nonumber\\
             &-\frac{31260191}{603741600}-\frac{5229\ln^2(3)}{170368}-\frac{10856743\ln(3)}{67082400}+\frac{3045587\ln(2)}{9583200}\Bigg]v^3+\mathcal{O}(v^4),\\
  \nu_O(v) = & \frac{1}{2}+\frac{5}{44}v+\frac{215}{26136}v^2+\Bigg[\frac{20}{121} \text{Li}_2\left(\frac{1}{3}\right)-\frac{5\pi^2}{726}-\frac{32905}{574992}+\frac{5\ln^2(3)}{121}\nonumber\\
             &-\frac{1085 \ln(3)}{11979}+\frac{2170\ln(2)}{11979}\Bigg]v^3+\mathcal{O}(v^4),\\
  \gamma_D(v) = & 1+\frac{5}{33}v+\frac{1}{99}v^2+\Bigg[\frac{11021 \text{Li}_2\left(\frac{1}{3}\right)}{85184}-\frac{377\pi^2}{92928}-\frac{130211}{1341648}+\frac{2919\ln^2(3)}{85184}\nonumber\\
             &+\frac{138365\ln(3)}{1341648}-\frac{7631\ln(2)}{87120}\Bigg]v^3+\mathcal{O}(v^4),\qquad\\
  \gamma_O(v) = & 1+\frac{5}{22}v+\frac{5}{484}v^2+\Bigg[\frac{120}{121} \text{Li}_2\left(\frac{1}{3}\right)-\frac{5 \pi^2}{121}-\frac{19315}{95832}+\frac{30 \ln^2(3)}{121}\nonumber\\
             &+\frac{150 \ln (3)}{1331}-\frac{300\ln(2)}{1331}\Bigg]v^3+\mathcal{O}(v^4).
\end{align}
\endgroup
In the case of SSM, the coefficients of these expansions, written in decimals, completely coincide with those presented in Ref.~\cite{NickelPRB1976,PhysRevB.17.1365}.

\section{Individual diagrams results}
Below we provide the results organized in the form of tables with explicit
contributions of calculation. For each diagram we provide its unique id,
perturbative order, graph structure encoded via Nickel notation and figure of
its edge structure. Finally we provide analytical expression, which for the SSM
case corresponds to numerical values provided in \cite{NickelREPORT}.

\subsection{2-point functions}
\label{sec:dias-2pt}
\begin{center}
  \begin{tabularx}{\textwidth}{p{1cm}p{1cm}p{3cm}c}
    \hline
    1 & $W^0$ &\texttt{//} &
                             $\frac{\partial}{\partial q^2}\vcenter{\hbox{
                             \begin{tikzpicture}[use Hobby shortcut, scale=0.8]
                               \draw (-0.75,0) -- (0.75,0);
                               \path[use as bounding box] (-1,-0.8) rectangle (1,0.8);
                             \end{tikzpicture}
                             }}$
    \\\hline
  \end{tabularx}
\end{center}

\begin{align}
  \Gamma_{O,1}^{(2,0)} & = 1\label{eq:dia1-2-N}\\
  \Gamma_{D,1}^{(2,0)} & = 1\label{eq:dia1-2-D}
\end{align}

\begin{center}
  \begin{tabularx}{\textwidth}{p{1cm}p{1cm}p{3cm}c}
    \hline
    2 & $W^2$ &\texttt{/E111/E//} &
                                    $\frac{\partial}{\partial q^2}\vcenter{\hbox{
                                    \begin{tikzpicture}[use Hobby shortcut, scale=0.8]
                                      \draw (-0.65,0) .. (0.0,0.35) .. (0.65,0);
                                      \draw (-0.65,0) .. (0.0,-0.35) .. (0.65,0);
                                      \draw (-0.75,0) -- (0.75,0);
                                      \fill (-0.65,0) circle (1pt);
                                      \fill (0.65,0) circle (1pt);
                                      \path[use as bounding box] (-1,-0.8) rectangle (1,0.8);
                                    \end{tikzpicture}
                                    }}$
    \\\hline
  \end{tabularx}
\end{center}

\begin{align}
  \Gamma_{O,1}^{(2,0)} & = \frac{4}{27} (n + 2)\label{eq:dia2-2-N}\\
  \Gamma_{D,1}^{(2,0)} & = \frac{2}{135} (-181 + 792 \ln{2} - 315 \ln{3})\label{eq:dia2-2-D}
\end{align}

\begin{center}
  \begin{tabularx}{\textwidth}{p{1cm}p{1cm}p{3cm}c}
    \hline
    3 & $W^3$ &\texttt{/E112/22/E/} &
                                      $\frac{\partial}{\partial q^2}\vcenter{\hbox{
                                      \begin{tikzpicture}[use Hobby shortcut, scale=0.8]
                                        \draw (-0.65,0) .. (-0.3,0.15) .. (0,0);
                                        \draw (-0.65,0) .. (-0.3,-0.15) .. (0,0);
                                        \draw (0,0) .. (0.3,0.15) .. (0.65,0);
                                        \draw (0,0) .. (0.3,-0.15) .. (0.65,0);
                                        \draw (-0.65,0) .. (0.0,-0.35) .. (0.65,0);
                                        \draw (-0.75,0) -- (-0.65,0);
                                        \draw (0.65,0) -- (0.75,0);
                                        \fill (-0.65,0) circle (1pt);
                                        \fill (0,0) circle (1pt);
                                        \fill (0.65,0) circle (1pt);
                                        \path[use as bounding box] (-1,-0.8) rectangle (1,0.8);
                                      \end{tikzpicture}
                                      }}$
    \\\hline
  \end{tabularx}
\end{center}

\begin{align}
  \Gamma_{O,2}^{(2,0)} & = -\frac{4}{27} (2 + n) (8 + n) \left(8 + 3 \pi^2 + 64 \ln{2} - 
                         32 \ln{3} - 18 \ln^2{3} - 72 \Li{2}\left( \frac{1}{3} \right)\right)\label{eq:dia3-2-N}\\
  \Gamma_{D,2}^{(2,0)} & = -\frac{1}{21600}\left(464880 + 503595 \pi^2 + 11281712 \ln{2} - 
                         7154576 \ln{3} \right.\nonumber\\
                       & \left.- 1833300 \ln^2{3} - 
                         8843580 \Li{2}\left( \frac{1}{3} \right)\right)\label{eq:dia3-2-D}
\end{align}

\subsection{4-point functions}
\label{sec:dias-4pt}

\begin{center}
  \begin{tabularx}{\textwidth}{p{1cm}p{1cm}p{3cm}c}
    \hline
    1 & $W^0$ & \texttt{/EEEE/} & $\vcenter{\hbox{
                                  \begin{tikzpicture}[use Hobby shortcut, scale=0.8]
                                    \draw (45:0.3) -- (0,0); \draw (-45:0.3) -- (0,0); \draw (135:0.3) --
                                    (0,0); \draw (-135:0.3) -- (0,0); \fill (0,0) circle (1pt);
                                    \path[use as bounding box] (-1,-0.8) rectangle (1,0.8);
                                  \end{tikzpicture}
                                  }}$\\
    \hline
  \end{tabularx}
\end{center}

\begin{align}
  \Gamma_{O,1}^{(4,0)} & = 1\label{eq:dia1-4-N}\\
  \Gamma_{D,1}^{(4,0)} & = 1\label{eq:dia1-4-D}\\
  \Gamma_{O,1}^{(2,1)} & = 1\label{eq:dia1-21-N}\\
  \Gamma_{D,1}^{(2,1)} & = 1\label{eq:dia1-21-D}                          
\end{align}

\begin{center}
  \begin{tabularx}{\textwidth}{p{1cm}p{1cm}p{3cm}c}
    \hline
    2 & $W^1$ & \texttt{/EE11/EE/} &
                                     $\vcenter{\hbox{
                                     \begin{tikzpicture}[use Hobby shortcut, scale=0.8]
                                       \draw (-0.35,-0.1) .. (90:0.15) .. (0.35,-0.1);
                                       \draw (-0.35,0.1) .. (-90:0.15) .. (0.35,0.1);
                                       \fill (-0.3,0) circle (1pt);
                                       \fill (0.3,0) circle (1pt);
                                       \path[use as bounding box] (-1,-0.8) rectangle (1,0.8);
                                     \end{tikzpicture}
                                     }}$
    \\\hline
  \end{tabularx}
\end{center}

\begin{align}
  \Gamma_{O,2}^{(4,0)} & = (n+8) \label{eq:dia2-4-N}\\
  \Gamma_{D,2}^{(4,0)} & = \frac{34}{5}\label{eq:dia2-4-D}\\
  \Gamma_{O,2}^{(2,1)} & = (n+2)\label{eq:dia2-21-N}\\
  \Gamma_{D,2}^{(2,1)} & = \frac{10}{3}\label{eq:dia2-21-D}                          
\end{align}

\begin{center}
  \begin{tabularx}{\textwidth}{p{1cm}p{1cm}p{3cm}c}
    \hline
    3 & $W^2$ &\texttt{/EE11/22/EE/} &
                                       $\vcenter{\hbox{
                                       \begin{tikzpicture}[use Hobby shortcut, scale=0.8]
                                         \draw (-0.7,-0.1) .. (-0.3,0.15) .. (0,0);
                                         \draw (-0.7,0.1) .. (-0.3,-0.15) .. (0,0);
                                         \draw (0,0) .. (0.3,0.15) .. (0.7,-0.1);
                                         \draw (0,0) .. (0.3,-0.15) .. (0.7,0.1);
                                         \fill (-0.65,0) circle (1pt);
                                         \fill (0,0) circle (1pt);
                                         \fill (0.65,0) circle (1pt);
                                         \path[use as bounding box] (-1,-0.8) rectangle (1,0.8);
                                       \end{tikzpicture}
                                       }}$
    \\\hline
  \end{tabularx}
\end{center}

\begin{align}
  \Gamma_{O,3}^{(4,0)} & = (n^2+6n +20) \label{eq:dia3-4-N}\\
  \Gamma_{D,3}^{(4,0)} & = \frac{492}{25}\label{eq:dia3-4-D}\\
  \Gamma_{O,3}^{(2,1)} & = (n+2)^2\label{eq:dia3-21-N}\\
  \Gamma_{D,3}^{(2,1)} & = \frac{100}{9}\label{eq:dia3-21-D}                          
\end{align}

\begin{center}
  \begin{tabularx}{\textwidth}{p{1cm}p{1cm}p{3cm}c}
    \hline
    4 & $W^2$ &\texttt{/EE12/E22/E/} &
                                       $\vcenter{\hbox{
                                       \begin{tikzpicture}[use Hobby shortcut, scale=0.8]
                                         \draw (-0.7,-0.1) .. (-0.3,0.15) .. (0.3,0.35);
                                         \draw (-0.7,0.1) .. (-0.3,-0.15) .. (0.3,-0.35);
                                         \draw (0.3,0.35) .. (0.2,0) .. (0.3,-0.35);
                                         \draw (0.3,0.35) .. (0.4,0) .. (0.3,-0.35);
                                         \draw (0.3,0.35) -- (0.4,0.45);
                                         \draw (0.3,-0.35) -- (0.4,-0.45);
                                         \fill (-0.57,0) circle (1pt);
                                         \fill (0.3,0.35) circle (1pt);
                                         \fill (0.3,-0.35) circle (1pt);
                                         \path[use as bounding box] (-1,-0.8) rectangle (1,0.8);
                                       \end{tikzpicture}
                                       }}$
    \\\hline
  \end{tabularx}
\end{center}

\begin{align}
  \Gamma_{O,4}^{(4,0)} & = \frac{8}{3}(5n +22) \label{eq:dia4-4-N}\\
  \Gamma_{D,4}^{(4,0)} & = \frac{1612}{45} - \frac{32}{15}\ln{3}\label{eq:dia4-4-D}\\
  \Gamma_{O,4}^{(2,1)} & = 4(n+2)\label{eq:dia4-21-N}\\
  \Gamma_{D,4}^{(2,1)} & = \frac{22}{3}\label{eq:dia4-21-D}                          
\end{align}

\begin{center}
  \begin{tabularx}{\textwidth}{p{1cm}p{1cm}p{3cm}c}
    \hline
    5 & $W^3$ & \texttt{/EE11/22/33/EE/} &
                                           $\vcenter{\hbox{
                                           \begin{tikzpicture}[use Hobby shortcut, scale=0.8]
                                             \draw (-1.0,-0.1) .. (-0.6,0.15) .. (-0.3,0);
                                             \draw (-1.0,0.1) .. (-0.6,-0.15) .. (-0.3,0);
                                             \draw (0.3,0) .. (0.6,0.15) .. (1.0,-0.1);
                                             \draw (0.3,0) .. (0.6,-0.15) .. (1.0,0.1);
                                             \draw (-0.3,0) .. (0.0,0.15) .. (0.3,0);
                                             \draw (-0.3,0) .. (0.0,-0.15) .. (0.3,0);
                                             \fill (-0.95,0) circle (1pt);
                                             \fill (-0.3,0) circle (1pt);
                                             \fill (0.3,0) circle (1pt);
                                             \fill (0.95,0) circle (1pt);
                                             \path[use as bounding box] (-1,-0.8) rectangle (1,0.8);
                                           \end{tikzpicture}
                                           }}$
    \\\hline
  \end{tabularx}
\end{center}

\begin{align}
  \Gamma_{O,5}^{(4,0)} & = (n^3 + 8 n^2 + 24 n + 48) \label{eq:dia5-4-N}\\
  \Gamma_{D,5}^{(4,0)} & = \frac{70664}{1125}\label{eq:dia5-4-D}\\
  \Gamma_{O,5}^{(2,1)} & = (n+2)^3\label{eq:dia5-21-N}\\
  \Gamma_{D,5}^{(2,1)} & = \frac{1000}{27}\label{eq:dia5-21-D}                          
\end{align}

\begin{center}
  \begin{tabularx}{\textwidth}{p{1cm}p{1cm}p{3cm}c}
    \hline
    6 &  $W^3$ & \texttt{/EE11/23/E33/E/} & 
                                            $\vcenter{\hbox{
                                            \begin{tikzpicture}[use Hobby shortcut, scale=0.8]
                                              \draw (-1.0,-0.1) .. (-0.6,0.15) .. (-0.3,0);
                                              \draw (-1.0,0.1) .. (-0.6,-0.15) .. (-0.3,0);
                                              \draw (-0.3,0) .. (0.0,0.2) .. (0.6,0.35);
                                              \draw (-0.3,0) .. (0.0,-0.2) .. (0.6,-0.35);
                                              \draw (0.6,0.35) .. (0.5,0) .. (0.6,-0.35);
                                              \draw (0.6,0.35) .. (0.7,0) .. (0.6,-0.35);
                                              \draw (0.6,0.35) -- (0.7,0.45);
                                              \draw (0.6,-0.35) -- (0.7,-0.45);
                                              \fill (-0.95,0) circle (1pt);
                                              \fill (-0.3,0) circle (1pt);
                                              \fill (0.6,0.35) circle (1pt);
                                              \fill (0.6,-0.35) circle (1pt);
                                              \path[use as bounding box] (-1,-0.8) rectangle (1,0.8);
                                            \end{tikzpicture}
                                            }}$
    \\\hline
  \end{tabularx}
\end{center}

\begin{align}
  \Gamma_{O,6}^{(4,0)} & = \frac{8}{3} (3n^2+ 22 n + 56) \label{eq:dia6-4-N}\\
  \Gamma_{D,6}^{(4,0)} & = \frac{62168}{675} - \frac{448}{225}\ln{3}\label{eq:dia6-4-D}\\
  \Gamma_{O,6}^{(2,1)} & = 4 (n+2)^2\label{eq:dia6-21-N}\\
  \Gamma_{D,6}^{(2,1)} & = \frac{220}{9}\label{eq:dia6-21-D}                          
\end{align}

\begin{center}
  \begin{tabularx}{\textwidth}{p{1cm}p{1cm}p{3cm}c}
    \hline
    7 & $W^3$ &\texttt{/EE12/EE3/333//} &
                                          $\vcenter
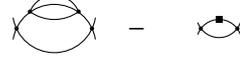
{\hbox{
                                          \begin{tikzpicture}[use Hobby shortcut, scale=0.8]
                                            \draw (195:0.7) .. (90:0.4) .. (-15:0.7);
                                            \draw (165:0.7) .. (-90:0.4) .. (15:0.7);
                                            \draw (-0.4,0.28) .. (90:0.55) .. (0.4,0.28);
                                            \draw (-0.4,0.28) .. (90:0.15) .. (0.4,0.28); 
                                            \fill (-0.4,0.28) circle (1pt);
                                            \fill (0.4,0.28) circle (1pt);
                                            \fill (0:0.62) circle (1pt);
                                            \fill (180:0.62) circle (1pt);
                                            \path[use as bounding box] (-1,-0.8) rectangle (1,0.8);
                                          \end{tikzpicture}
                                          }}
                                          -
                                          \vcenter{\hbox{
                                          \begin{tikzpicture}[use Hobby shortcut, scale=0.8]
                                            \draw (-0.35,-0.1) .. (90:0.15) .. (0.35,-0.1);
                                            \draw (-0.35,0.1) .. (-90:0.15) .. (0.35,0.1);
                                            \fill (-0.3,0) circle (1pt);
                                            \fill (0.3,0) circle (1pt);
                                            \fill (-0.05,0.10) rectangle (0.05,0.2);
                                            \path[use as bounding box] (-1,-0.8) rectangle (1,0.8);
                                          \end{tikzpicture}
                                          }}
                                          $
    \\\hline
  \end{tabularx}
\end{center}

\begin{align}
  \Gamma_{O,7}^{(4,0)} & = (n+8)(n+2)(1-8\ln{2}+4\ln{3}) \label{eq:dia7-4-N}\\
  \Gamma_{D,7}^{(4,0)} & = \frac{294916}{1575} + \frac{390592}{225} \ln{2} - \frac{665516}{525}\ln{3}\label{eq:dia7-4-D}\\
  \Gamma_{O,7}^{(2,1)} & = (n+2)^2 (1-8\ln{2}+4\ln{3})\label{eq:dia7-21-N}\\
  \Gamma_{D,7}^{(2,1)} & = \frac{17348}{189} + \frac{22976}{27} \ln{2} - 
                         \frac{39148}{63} \ln{3}\label{eq:dia7-21-D}                          
\end{align}

\begin{center}
  \begin{tabularx}{\textwidth}{p{1cm}p{1cm}p{3cm}c}
    \hline
    8 & $W^3$ &\texttt{/EE12/E23/33/E/} &
                                          $\vcenter
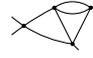
{\hbox{
                                          \begin{tikzpicture}[use Hobby shortcut, scale=0.8]
                                            \draw (0,0.3) .. (0.3,0.4) .. (0.6,0.3);
                                            \draw (0,0.3) .. (0.3,0.2) .. (0.6,0.3);
                                            \draw (-0.7,-0.15) .. (-0.5,0) .. (0,0.3);
                                            \draw (-0.7,0.15) .. (-0.5,0) .. (0.3,-0.3);
                                            \draw (0.3,-0.3) -- (0,0.3);
                                            \draw (0.3,-0.3) -- (0.6,0.3);
                                            \draw (0.3,-0.3) -- (0.4,-0.4);
                                            \draw (0.6,0.3) -- (0.7,0.35);
                                            \fill (-0.5,0) circle (1pt);
                                            \fill (0,0.3) circle (1pt);
                                            \fill (0.6,0.3) circle (1pt);
                                            \fill (0.3,-0.3) circle (1pt);
                                            \path[use as bounding box] (-1,-0.8) rectangle (1,0.8);
                                          \end{tikzpicture}
                                          }}$
    \\\hline
  \end{tabularx}
\end{center}

\begin{align}
  \Gamma_{O,8}^{(4,0)} & = \frac{64}{3}(n^2+ 20 n + 60)(2\ln{2}-\ln{3}) \label{eq:dia8-4-N}\\
  \Gamma_{D,8}^{(4,0)} & = \frac{1}{4725}\left(-1471408 + 113190 \pi^2 - 9058352 \ln{2} + 
                         6623724 \ln{3} \right.\nonumber\\
                       & \left.+ 91665 \ln^2{3} - 104580 \Li{2}\left( \frac{1}{3} \right)\right)
                         \label{eq:dia8-4-D}\\
  \Gamma_{O,8}^{(2,1)} & = \frac{32}{3}(n+8)(n+2)(2\ln{2}-\ln{3})\label{eq:dia8-21-N}\\
  \Gamma_{D,8}^{(2,1)} & = \frac{1}{5040} \left(29855 \pi^2 - 
                         4 (89620 + 354212 \ln{2} \right.\nonumber\\
                       & \left.- \ln{3} (276596 + 4515 \ln{3}) + 
                         11235 \Li{2}\left( \frac{1}{3}\right)\right)
                         \label{eq:dia8-21-D}                          
\end{align}

\begin{center}
  \begin{tabularx}{\textwidth}{p{1cm}p{1cm}p{3cm}c}
    \hline
    9 & $W^3$ &\texttt{/EE12/E33/E33//} &
                                          $\vcenter
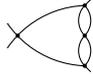
{\hbox{
                                          \begin{tikzpicture}[use Hobby shortcut, scale=0.8]
                                            \draw (0.5,0.5) -- (0.6,0.6);
                                            \draw (0.5,-0.5) -- (0.6,-0.6);
                                            \draw (0.5,0.5) .. (0.6,0.25) .. (0.5,0);
                                            \draw (0.5,0.5) .. (0.4,0.25) .. (0.5,0);
                                            \draw (0.5,-0.5) .. (0.6,-0.25) .. (0.5,0);
                                            \draw (0.5,-0.5) .. (0.4,-0.25) .. (0.5,0);
                                            \draw (195:0.8) .. (90:0.4) .. (0.5,0.5);
                                            \draw (165:0.8) .. (-90:0.4) .. (0.5,-0.5);
                                            \fill (0.5,0.5) circle (1pt);
                                            \fill (0.5,-0.5) circle (1pt);
                                            \fill (0.5,0) circle (1pt);
                                            \fill (-0.6,0) circle (1pt);
                                            \path[use as bounding box] (-1,-0.8) rectangle (1,0.8);
                                          \end{tikzpicture}
                                          }}$
    \\\hline
  \end{tabularx}
\end{center}

\begin{align}
  \Gamma_{O,9}^{(4,0)} & = -\frac{4}{3} (3n^2+22n +56) \left(\pi^2 + 32 \ln{2} - 
                         2 \ln{3} (8 + 3 \ln{3}) - 24 \Li{2}\left( \frac{1}{3} \right)\right)
                         \label{eq:dia9-4-N}\\
  \Gamma_{D,9}^{(4,0)} & = \frac{1}{302400} \left( -14660835 \pi^2 + 
                         4 \left(11249764 + 98429460 \ln{2} - 62066844 \ln{3} \right.\right.\nonumber\\
                       & \left.\left.+ 
                         10205685 \ln{3}^2 + 47160855 \Li{2}\left( \frac{1}{3} \right)\right)\right)
                         \label{eq:dia9-4-D}\\
  \Gamma_{O,9}^{(2,1)} & = -\frac{2}{3} (2 + n) (8 + n)  \left(\pi^2 + 32 \ln{2} - 
                         2 \ln{3} (8 + 3 \ln{3}) - 24 \Li{2}\left( \frac{1}{3} \right)\right)
                         \label{eq:dia9-21-N}\\
  \Gamma_{D,9}^{(2,1)} & = \frac{1}{40320} \left(-383985 \pi^2 + 
                         4 \left(297868 + 2593820 \ln{2} - 1642228 \ln{3} \right.\right.\nonumber\\
                       & \left.\left.+ 270375 \ln{3}^2 + 
                         1247085 \Li{2}\left( \frac{1}{3} \right)\right)\right)
                         \label{eq:dia9-21-D}                          
\end{align}

\begin{center}
  \begin{tabularx}{\textwidth}{p{1cm}p{1cm}p{3cm}c}
    \hline
    10 & $W^3$ & \texttt{/EE12/223/3/EE/} &
                                            $\vcenter{\hbox{
                                            \begin{tikzpicture}[use Hobby shortcut, scale=0.8]
                                              \draw (195:0.8) .. (90:0.4) .. (-15:0.8);
                                              \draw (165:0.8) .. (-90:0.4) .. (15:0.8);
                                              \draw (90:0.4) .. (-0.1,0) .. (-90:0.4);
                                              \draw (90:0.4) .. (0.1,0) .. (-90:0.4);
                                              \fill (180:0.7) circle (1pt);
                                              \fill (0:0.7) circle (1pt);
                                              \fill (90:0.4) circle (1pt);
                                              \fill (-90:0.4) circle (1pt);
                                              \path[use as bounding box] (-1,-0.8) rectangle (1,0.8);
                                            \end{tikzpicture}
                                            }}$
    \\\hline
  \end{tabularx}
\end{center}

\begin{align}
  \Gamma_{O,10}^{(4,0)} & =   (3 n^2 + 22 n +56)\label{eq:dia10-4-N}\\
  \Gamma_{D,10}^{(4,0)} & = -\frac{8}{2027025} (44913784 + 465443952 \ln{2} - 
                          342235935 \ln{3})\label{eq:dia10-4-D}\\
  \Gamma_{O,10}^{(2,1)} & = 3 (n + 2)^2\label{eq:dia10-21-N}\\
  \Gamma_{D,10}^{(2,1)} & = -\frac{8}{63} (536 + 7280 \ln{2} - 5211 \ln{3})\label{eq:dia10-21-D}
\end{align}

\begin{center}
  \begin{tabularx}{\textwidth}{p{1cm}p{1cm}p{3cm}c}
    \hline
    11 & $W^3$ & \texttt{/E112/E3/E33/E/} &
                                            $\vcenter{\hbox{
                                            \begin{tikzpicture}[use Hobby shortcut, scale=0.8]
                                              \draw (0,0) circle (0.5);
                                              \draw (45:0.7) .. (45:0.5) .. (-45:0.5) .. (-45:0.7);
                                              \draw (135:0.7) .. (135:0.5) .. (-135:0.5) .. (-135:0.7);
                                              \fill (45:0.5) circle (1pt);
                                              \fill (-45:0.5) circle (1pt);
                                              \fill (135:0.5) circle (1pt);
                                              \fill (-135:0.5) circle (1pt);
                                              \path[use as bounding box] (-1,-0.8) rectangle (1,0.8);
                                            \end{tikzpicture}
                                            }}$
    \\\hline
  \end{tabularx}
\end{center}

\begin{align}
  \Gamma_{O,11}^{(4,0)} & =   -\frac{4}{3}  (n^2 +20 n +60)  \left(\pi^2 + 32 \ln{2} - 
                          2 \ln{3} (8 + 3 \ln{3}) - 24 \Li{2}\left( \frac{1}{3} \right)\right)
                          \label{eq:dia11-4-N}\\
  \Gamma_{D,11}^{(4,0)} & = \frac{1}{10800} \left(384784 - 345315 \pi^2 + 5067600 \ln{2} - 
                          2426544 \ln{3} \right.\nonumber\\
                        & \left.+ 907740 \ln^2{3} + 4453020 \Li{2}\left( \frac{1}{3} \right)\right)
                          \label{eq:dia11-4-D}
\end{align}

\begin{center}
  \begin{tabularx}{\textwidth}{p{1cm}p{1cm}p{3cm}c}
    \hline
    12 & $W^3$ & \texttt{/E123/E23/E3/E/} &
                                            $\vcenter{\hbox{
                                            \begin{tikzpicture}[use Hobby shortcut, scale=0.8]
                                              \draw (0,0) circle (0.5);
                                              \draw (45:0.5) -- (45:0.7);
                                              \draw (-45:0.5) -- (-45:0.7);
                                              \draw (135:0.5) -- (135:0.7);
                                              \draw (-135:0.5) -- (-135:0.7);
                                              \draw (135:0.5) -- (-45:0.5);
                                              \draw (-135:0.5) -- ($(-135:0.5)!0.80!(0,0)$);
                                              \draw (45:0.5) -- ($(45:0.5)!0.80!(0,0)$);
                                              \fill (45:0.5) circle (1pt);
                                              \fill (-45:0.5) circle (1pt);
                                              \fill (135:0.5) circle (1pt);
                                              \fill (-135:0.5) circle (1pt);
                                              \path[use as bounding box] (-1,-0.8) rectangle (1,0.8);
                                            \end{tikzpicture}
                                            }}$
    \\\hline
  \end{tabularx}
\end{center}

\begin{align}
  \Gamma_{O,12}^{(4,0)} & = 32 \sqrt{2} (5n+22) \left( \CL{2}\left(4\alpha \right) - \CL{2}\left( 2\alpha \right) \right)  \label{eq:dia12-4-N}\\
  \Gamma_{D,12}^{(4,0)} & = \frac{1}{225} \left(
                          -9176
                          + 1290 \sqrt{2} \left( \CL{2}\left(4\alpha \right) - \CL{2}\left( 2\alpha \right) \right)
                          + 5 (-3497 + 822 \sqrt{2}) \pi^2
                          \right.\nonumber\\
                        & + 6165 \sqrt{2} \arccos^2{3} + 9 \ln{3} (-23976 - 36850 \ln{2} + 4605 \ln{3})
                          + 360440 \ln{2}\nonumber\\
                        & \left. + 568770 \ln^2{2} + 284385 \Li{2}\left( \frac{1}{4} \right)
                          - 24660 \sqrt{2} \Li{2}\left( 3-2\sqrt{2} \right)
                          \right)
                          \label{eq:dia12-4-D}
\end{align}
Where we have introduced angle $\alpha = \arcsin{\frac{1}{3}}$ according
to\cite{Broadhurst:1998iq} and Clausen function is defined as $\CL{2}(\phi) = \mathrm{Im}\,\left[ \Li{2}\left(e^{i\phi}\right) \right]$.

\bibliographystyle{JHEP}
\bibliography{ginf3l}

\providecommand{\href}[2]{#2}\begingroup\raggedright\begin{thebibliography}{10}

\bibitem{PhysRevB.4.3174}
K.~Wilson, \emph{Renormalization group and critical phenomena. {I}.
  {R}enormalization group and the {K}adanoff scaling picture}, {\emph{Phys.
  Rev. B} {\bfseries 4} (1971) 3174}.

\bibitem{PhysRevB.4.3184}
K.~Wilson, \emph{Renormalization group and critical phenomena. {II}.
  {P}hase-space cell analysis of critical behavior}, {\emph{Phys. Rev. B}
  {\bfseries 4} (1971) 3184}.

\bibitem{WILSON1974119}
K.~Wilson, \emph{Critical phenomena in 3.99 dimensions}, {\emph{Physica}
  {\bfseries 73} (1974) 119}.

\bibitem{PhysRevLett.28.240}
K.G.~Wilson and M.E.~Fisher, \emph{Critical exponents in 3.99 dimensions},
  \href{https://doi.org/10.1103/PhysRevLett.28.240}{\emph{Phys. Rev. Lett.}
  {\bfseries 28} (1972) 240}.

\bibitem{CHETYRKIN1983351}
K.G.~Chetyrkin, S.G.~Gorishny, S.A.~Larin and F.V.~Tkachov, \emph{Five-loop
  renormalization group calculations in the $g\varphi^4$ theory},
  \href{https://doi.org/https://doi.org/10.1016/0370-2693(83)90324-6}{\emph{Phys.
  Lett. B} {\bfseries 132} (1983) 351}.

\bibitem{CHETYRKIN1981147}
K.G.~Chetyrkin, A.L.~Kataev and F.V.~Tkachov, \emph{Five-loop calculations in
  the $g\varphi^4$ model and the critical index $\eta$},
  \href{https://doi.org/https://doi.org/10.1016/0370-2693(81)90968-0}{\emph{Phys.
  Lett. B} {\bfseries 99} (1981) 147}.

\bibitem{LeGuillou1985}
J.C.L.~Guillou and J.~Zinn-Justin, \emph{Accurate critical exponents from the
  $\varepsilon$-expansion},
  \href{https://doi.org/10.1051/jphyslet:01985004604013700}{\emph{Journal de
  Physique Lettres} {\bfseries 46} (1985) 137}.

\bibitem{KLEINERT199139}
H.~Kleinert, J.~Neu, N.~Schulte-Frohlinde, K.G.~Chetyrkin and S.A.~Larin,
  \emph{Five-loop renormalization group functions of $o(n)$-symmetric
  $\varphi$-theory and $\varepsilon$-expansions of critical exponents up to
  $\varepsilon^5$.},
  \href{https://doi.org/https://doi.org/10.1016/0370-2693(91)91009-K}{\emph{Phys.
  Lett. B} {\bfseries 272} (1991) 39}.

\bibitem{BinothHeinrich:SectorDecomposition}
T.~Binoth and G.~Heinrich, \emph{An automatized algorithm to compute infrared
  divergent multiloop integrals},
  \href{https://doi.org/10.1016/S0550-3213(00)00429-6}{\emph{Nucl. Phys. B}
  {\bfseries 585} (2000) 741}.

\bibitem{Baikov2010mi}
P.A.~Baikov and K.G.~Chetyrkin, \emph{Four loop massless propagators: An
  algebraic evaluation of all master integrals},
  \href{https://doi.org/10.1016/j.nuclphysb.2010.05.004}{\emph{Nucl. Phys. B}
  {\bfseries 837} (2010) 186}.

\bibitem{Lee2012mi}
R.N.~Lee, A.V.~Smirnov and V.A.~Smirnov, \emph{Master integrals for four-loop
  massless propagators up to weight twelve},
  \href{https://doi.org/10.1016/j.nuclphysb.2011.11.005}{\emph{Nucl. Phys. B}
  {\bfseries 856} (2012) 95}.

\bibitem{chetyrkin2017rstar}
K.G.~Chetyrkin, \emph{Combinatorics of $\bf{R}-$, $\bf{R^{-1}}$, and
  $\bf{R^{*}}$-operations and asymptotic expansions of feynman integrals in the
  limit of large momenta and masses}, {\emph{arXiv preprint arXiv:1701.08627}
  (2017) }.

\bibitem{Batkovich2015rstar}
D.V.~Batkovich and M.V.~Kompaniets, \emph{Toolbox for multiloop feynman
  diagrams calculations using $r^*$ operation},
  \href{https://doi.org/10.1088/1742-6596/608/1/012068}{\emph{J. Phys. Conf.
  Ser.} {\bfseries 608} (2015) 012068}.

\bibitem{Brown:TwoPoint}
F.C.S.~Brown, \emph{The massless higher-loop two-point function},
  \href{https://doi.org/10.1007/s00220-009-0740-5}{\emph{Commun. Math. Phys.}
  {\bfseries 287} (2009) 925}.

\bibitem{Panzer:HyperIntAlgorithms}
E.~Panzer, \emph{Algorithms for the symbolic integration of hyperlogarithms
  with applications to {Feynman} integrals},
  \href{https://doi.org/10.1016/j.cpc.2014.10.019}{\emph{Comp. Phys. Commun.}
  {\bfseries 188} (2015) 148}.

\bibitem{BrownKreimer:AnglesScales}
F.C.S.~Brown and D.~Kreimer, \emph{Angles, scales and parametric
  renormalization},
  \href{https://doi.org/10.1007/s11005-013-0625-6}{\emph{Lett. Math. Phys.}
  {\bfseries 103} (2013) 933}.

\bibitem{BatkovichKompanietsChetyrkin:6loop}
D.V.~Batkovich, M.V.~Kompaniets and K.G.~Chetyrkin, \emph{Six loop analytical
  calculation of the field anomalous dimension and the critical exponent
  {$\eta$} in {$O(n)$}-symmetric {$\varphi^4$} model},
  \href{https://doi.org/10.1016/j.nuclphysb.2016.03.009}{\emph{Nucl. Phys. B}
  {\bfseries 906} (2016) 147}.

\bibitem{KompanietsPanzer:LL2016}
M.V.~Kompaniets and E.~Panzer, \emph{Renormalization group functions of
  {$\phi^4$} theory in the {MS}-scheme to six loops},  in \emph{Loops and Legs
  in Quantum Field Theory Leipzig, Germany, April 24--29}, Proceedings of
  Science, 2016,
  \href{http://pos.sissa.it/cgi-bin/reader/contribution.cgi?id=260/038}{http://pos.sissa.it/cgi-bin/reader/contribution.cgi?id=260/038}
  [\href{https://arxiv.org/abs/1606.09210}{{\ttfamily 1606.09210}}].

\bibitem{KP17}
M.V.~Kompaniets and E.~Panzer, \emph{Minimally subtracted six-loop
  renormalization of $o(n)$-symmetric $\phi$ 4 theory and critical exponents},
  {\emph{Phys. Rev. D} {\bfseries 96} (2017) 036016}.

\bibitem{PhysRevD.97.085018}
O.~Schnetz, \emph{Numbers and functions in quantum field theory},
  \href{https://doi.org/10.1103/PhysRevD.97.085018}{\emph{Phys. Rev. D}
  {\bfseries 97} (2018) 085018}.

\bibitem{six_loop_cubic}
L.T.~Adzhemyan, E.V.~Ivanova, M.V.~Kompaniets, A.~Kudlis and A.I.~Sokolov,
  \emph{Six-loop $\varepsilon$ expansion study of three-dimensional $n$-vector
  model with cubic anisotropy},
  \href{https://doi.org/https://doi.org/10.1016/j.nuclphysb.2019.02.001}{\emph{Nucl.
  Phys. B} {\bfseries 940} (2019) 332}.

\bibitem{six_loop_onom}
M.V.~Kompaniets, A.~Kudlis and A.I.~Sokolov, \emph{Six-loop $\varepsilon$
  expansion study of three-dimensional $o(n)\times o(m)$ spin models},
  \href{https://doi.org/https://doi.org/10.1016/j.nuclphysb.2019.114874}{\emph{Nucl.
  Phys. B} {\bfseries 950} (2020) 114874}.

\bibitem{six_loop_unum}
L.T.~Adzhemyan, E.V.~Ivanova, M.V.~Kompaniets, A.~Kudlis and A.I.~Sokolov,
  \emph{Six-loop $\varepsilon$ expansion of three-dimensional $u(n)\times u(m)$
  models},
  \href{https://doi.org/https://doi.org/10.1016/j.nuclphysb.2022.115680}{\emph{Nucl.
  Phys. B} {\bfseries 975} (2022) 115680}.

\bibitem{Bednyakov2021}
A.~Bednyakov and A.~Pikelner, \emph{Six-loop beta functions in general scalar
  theory}, \href{https://doi.org/10.1007/jhep04(2021)233}{\emph{J. High Energy
  Phys.} {\bfseries 2021} (2021) }.

\bibitem{NickelREPORT}
B.~Nickel, D.I.~Meiron and G.~Baker, ``University of guelph report.'' 1977.

\bibitem{NickelPRB1976}
G.~Baker, B.~Nickel, M.S.~Green and D.I.~Meiron{\emph{Phys. Rev. Lett.}
  {\bfseries 36} (1976) 1351}.

\bibitem{PhysRevB.21.3976}
J.C.~Le~Guillou and J.~Zinn-Justin, \emph{Critical exponents from field
  theory}, \href{https://doi.org/10.1103/PhysRevB.21.3976}{\emph{Phys. Rev. B}
  {\bfseries 21} (1980) 3976}.

\bibitem{PhysRevLett.39.95}
J.C.~Le~Guillou and J.~Zinn-Justin, \emph{Critical exponents for the $n$-vector
  model in three dimensions from field theory},
  \href{https://doi.org/10.1103/PhysRevLett.39.95}{\emph{Phys. Rev. Lett.}
  {\bfseries 39} (1977) 95}.

\bibitem{PhysRevLett.30.559}
M.E.~Fisher and A.~Aharony, \emph{Dipolar interactions at ferromagnetic
  critical points},
  \href{https://doi.org/10.1103/PhysRevLett.30.559}{\emph{Phys. Rev. Lett.}
  {\bfseries 30} (1973) 559}.

\bibitem{PhysRevB.10.2078}
A.D.~Bruce and A.~Aharony, \emph{Critical exponents of ferromagnets with
  dipolar interactions: Second-order $\ensuremath{\epsilon}$ expansion},
  \href{https://doi.org/10.1103/PhysRevB.10.2078}{\emph{Phys. Rev. B}
  {\bfseries 10} (1974) 2078}.

\bibitem{Korzhenevskii1978}
A.L.~Korzhenevskii and A.I.~Sokolov, \emph{{Critical fluctuations and splitting
  of phase transitioning a tetragonal ferroelectric}}, {\emph{JETP Lett.}
  {\bfseries 27} (1978) 239}.

\bibitem{9168923975ca4bcebb5c0a2bc92ab3b5}
A.~Sokolov and A.~Tagantsev, \emph{Phase transitions in a cubic crystal with
  dipolar forces and an anisotropic correlation function.}, {\emph{Journal of
  Experimental and Theoretical Physics} {\bfseries 49} (1979) 92}.

\bibitem{KORNBLIT1973531}
A.~Kornblit, G.~Ahlers and E.~Buehler, \emph{Heat capacity of rbmnf3 and euo
  near the magnetic phase transitions},
  \href{https://doi.org/https://doi.org/10.1016/0375-9601(73)90027-3}{\emph{Phys.
  Lett. A} {\bfseries 43} (1973) 531}.

\bibitem{RevModPhys.46.597}
M.E.~Fisher, \emph{The renormalization group in the theory of critical
  behavior}, \href{https://doi.org/10.1103/RevModPhys.46.597}{\emph{Rev. Mod.
  Phys.} {\bfseries 46} (1974) 597}.

\bibitem{RevModPhys.52.489}
G.~Ahlers, \emph{Critical phenomena at low temperature},
  \href{https://doi.org/10.1103/RevModPhys.52.489}{\emph{Rev. Mod. Phys.}
  {\bfseries 52} (1980) 489}.

\bibitem{WACHTER1979507}
P.~Wachter, \emph{Chapter 19 europium chalcogenides: Euo, eus, euse and eute},
  in \emph{Alloys and Intermetallics}, vol.~2 of \emph{Handbook on the Physics
  and Chemistry of Rare Earths}, p.~507 (1979),
  \href{https://doi.org/https://doi.org/10.1016/S0168-1273(79)02010-9}{DOI}.

\bibitem{PhysRevB.17.1365}
G.A.~Baker, B.G.~Nickel and D.I.~Meiron, \emph{Critical indices from
  perturbation analysis of the callan-symanzik equation},
  \href{https://doi.org/10.1103/PhysRevB.17.1365}{\emph{Phys. Rev. B}
  {\bfseries 17} (1978) 1365}.

\bibitem{Rajantie:1996np}
A.K.~Rajantie, \emph{{Feynman diagrams to three loops in three-dimensional
  field theory}},
  \href{https://doi.org/10.1016/S0550-3213(96)00474-9}{\emph{Nucl. Phys. B}
  {\bfseries 480} (1996) 729}
  [\href{https://arxiv.org/abs/hep-ph/9606216}{{\ttfamily hep-ph/9606216}}].

\bibitem{Tentyukov:1999is}
M.~Tentyukov and J.~Fleischer, \emph{{A Feynman diagram analyzer DIANA}},
  \href{https://doi.org/10.1016/S0010-4655(00)00147-8}{\emph{Comput. Phys.
  Commun.} {\bfseries 132} (2000) 124}
  [\href{https://arxiv.org/abs/hep-ph/9904258}{{\ttfamily hep-ph/9904258}}].

\bibitem{Nogueira:1991ex}
P.~Nogueira, \emph{{Automatic Feynman graph generation}},
  \href{https://doi.org/10.1006/jcph.1993.1074}{\emph{J. Comput. Phys.}
  {\bfseries 105} (1993) 279}.

\bibitem{Smirnov:2019qkx}
A.V.~Smirnov and F.S.~Chuharev, \emph{{FIRE6: Feynman Integral REduction with
  Modular Arithmetic}},
  \href{https://doi.org/10.1016/j.cpc.2019.106877}{\emph{Comput. Phys. Commun.}
  {\bfseries 247 } (2020) 106877}
  [\href{https://arxiv.org/abs/1901.07808}{{\ttfamily 1901.07808}}].

\bibitem{Lee:2012cn}
R.N.~Lee, \emph{{Presenting LiteRed: a tool for the Loop InTEgrals REDuction}},
   \href{https://arxiv.org/abs/1212.2685}{{\ttfamily 1212.2685}}.

\bibitem{Broadhurst:1998iq}
D.J.~Broadhurst, \emph{{A Dilogarithmic three-dimensional Ising tetrahedron}},
  \href{https://doi.org/10.1007/s100529900983}{\emph{Eur. Phys. J. C}
  {\bfseries 8} (1999) 363}
  [\href{https://arxiv.org/abs/hep-th/9805025}{{\ttfamily hep-th/9805025}}].

\bibitem{Broadhurst:1998ke}
D.J.~Broadhurst, \emph{{Solving differential equations for three loop diagrams:
  Relation to hyperbolic geometry and knot theory}},
  \href{https://arxiv.org/abs/hep-th/9806174}{{\ttfamily hep-th/9806174}}.

\bibitem{Lee:2010hs}
R.N.~Lee and I.S.~Terekhov, \emph{{Application of the DRA method to the
  calculation of the four-loop QED-type tadpoles}},
  \href{https://doi.org/10.1007/JHEP01(2011)068}{\emph{J. High Energy Phys.}
  {\bfseries 01} (2011) 068} [\href{https://arxiv.org/abs/1010.6117}{{\ttfamily
  1010.6117}}].

\bibitem{Davydychev:2000na}
A.I.~Davydychev and M.Y.~Kalmykov, \emph{{New results for the epsilon expansion
  of certain one, two and three loop Feynman diagrams}},
  \href{https://doi.org/10.1016/S0550-3213(01)00095-5}{\emph{Nucl. Phys. B}
  {\bfseries 605} (2001) 266}
  [\href{https://arxiv.org/abs/hep-th/0012189}{{\ttfamily hep-th/0012189}}].

\bibitem{KUDLIS2020114881}
A.~Kudlis and A.I.~Sokolov, \emph{Universal effective couplings of the
  three-dimensional n-vector model and field theory},
  \href{https://doi.org/https://doi.org/10.1016/j.nuclphysb.2019.114881}{\emph{Nucl.
  Phys. B} {\bfseries 950} (2020) 114881}.

\end{thebibliography}\endgroup
\end{document}